\documentclass[conference]{IEEEtran}

\ifCLASSINFOpdf
  \usepackage[pdftex]{graphicx}
  \DeclareGraphicsExtensions{.pdf,.jpeg,.png}
\else
\fi

\usepackage[cmex10]{amsmath}

\usepackage{url}

\begin{document}
\title {Taxonomy driven indicator scoring in MISP threat intelligence platforms}

\author{\IEEEauthorblockN{Sami Mokaddem\\ G\'erard Wagener  \\ Alexandre Dulaunoy  and \\ Andras Iklody}
\IEEEauthorblockA{\\
CIRCL- Computer Incident Response Center Luxembourg \\
Email: info@circl.lu\\
}
\and
\IEEEauthorblockN{\\ Cynthia Wagner}\\
\\
\IEEEauthorblockA{Fondation RESTENA\\
CSIRT\\
Email: cynthia.wagner@restena.lu}}


\maketitle

\begin{abstract}

IT security community is recently facing a change of trend from closed to open
working groups and from restrictive information to full information disclosure
and sharing. One major feature for this trend change is the number of incidents
and various Indicators of compromise (IoC) that appear on a daily base, which
can only be faced and solved in a collaborative way. Sharing information is
key to stay on top of the threats.

To cover the needs of having a medium for information sharing, different
initiatives were taken such as the Open Source Threat Intelligence Platform
called MISP \cite{misp-article}. At current state, this sharing and collection
platform has become far more than a malware information sharing platform.
It includes all kind of IoCs, malware and vulnerabilities, but also
financial threat or fraud information. Hence, the volume of information is
increasing and evolving.

In this paper we present implemented distributed data interaction methods for
MISP followed by a generic scoring model for decaying information that is
shared within MISP communities. As the MISP community members do not
have the same objectives, use cases and implementations of the scoring model
are discussed. A commonly encountered use case in practice is the detection of
indicators of compromise in operational networks. 
\end{abstract}


%
\IEEEpeerreviewmaketitle

\section*{Practical Experience Report}

\section{Introduction}

On a daily basis new threats appear and disappear on the scenes of cybercrime
with no indication that this phenomenon ceases shortly. Fighting threats as a
singular has become impossible nowadays and communities have formed to share 
information about threats and work collaboratively to handle the problems.

Collaboration and information sharing have become a key element in the world
of threat processing in incident response. Surely, on one side, sharing
information is a critical  point due to sensitive data it may include
respectively the authenticity of information itself, but on the other,
joint-efforts to handle a problem have direct impact on reaction time and
resources. The sudden appearance of different kind of information sharing
platforms over recent past confirmed this trend.

\newpage

In this paper, a scoring model for the open source threat intelligence
platform called MISP \cite{misp-article} is presented. The aim of
MISP permits various actors, be it from  private or public IT-communities to
share their information, IoCs, malware and other existing threats. MISP
is a peer to peer sharing platform. It is not unusual that a piece of
information transits through multiple nodes from the producer to the consumer,
raising trust and data quality issues as it is not always known how the
shared information was acquired and trusted. MISP provides various
features to share additional information about its context. Currently, 47
MISP taxonomies are available for providing some context to a piece of
information which are regularly used in MISP communities.

A lot of research activities have been done in the domain of introduction 
systems ranging from the automated derivation of signatures to the exploration 
of anomaly detection techniques to the application of machine learning. 
However, only little effort was done to efficiently share signatures for intrusion 
detection system or even care about their validity or freshness. Since 
its early beginning, MISP was capable of sharing and exporting IDS signatures, 
which are ingested by intrusion detection systems. It also provides features where
intrusion detection systems or humans can offer feedback, whether they have seen a 
given piece of information, or if it is a false positive or whether it will expire soon.
This feedback is then distributed in the communities following 
the peer to peer data distribution model of MISP.

This paper presents implemented data interaction models in MISP, proposes the scoring model 
that uses these operational parameters to help to consume attributes and take decision about them. 
The paper is organized as follows,  section \ref{related} discusses relevant approaches focussing 
on threat intelligence collection, processing and sharing. 
Section \ref{misp-tool} is a general introduction to the MISP plateform. 
In section \ref{dint}, the data interaction methods are presented. 
In section \ref{sighting} attribute scoring methods are introduced to the reader. 
Since research is still ongoing, some future work and conclusions are presented in section \ref{conclusions}.

\newpage

\section{Related work}
\label{related}

Recent studies in cybersecurity \cite{Enisa}\cite{Haass2015} showed that one
major key to successful cyber incident response is information sharing in its
different forms, either by trusted third parties, email lists of CERTs
(Computer Emergency Response Teams) or threat intelligence and information
sharing platforms.

There are many information sharing platforms available in the IT world,
whereas a lot of them are commercial enterprises selling threat intelligence
as a service in their portfolios.

Besides these commercial tools, a lot of effort has already been made in the
 open domain too. Different initiatives such as STIX/TAXII
\cite{taxii}\cite{Stix}  have been developed to set up specifications for
cyber threat information exchange and standardized communication languages for information sharing.

The platform we used for implementing the scoring method is MISP. In
\cite{misp-article}, the MISP platform is described from a technical
 perspective and its main modules introduced. A major advantage is that MISP
is a collaborative open-source project that continuously evolves by
community-driven effort. Each member can consume or produce threat intelligence data.

A case study on information sharing is presented in \cite{Haass2015},
where a survey was performed on different aspects as hurdles, problems and
legal aspects in information sharing. The major outcome was that information
sharing remains a group or community activity. Another interesting outcome of
this survey was the need for accurate information sharing practices and low
false positive rates.

Information sharing is related to a lot of challenges as presented in
\cite{Brown2015}, where requirements and needs for successful threat
intelligence platforms, such as the added value of shared data and privacy,
are discussed. Other requirements are for example a new quality control approaches.

In \cite{Dandurand13}, the facility to share information, automate
information sharing and the ability to generate, refine and control data is
discussed by redefining these problems into a concept of knowledge management for the area of cyber security by adding user needs. This leads to the fact of the veracity of information and also to false positives. In
\cite{Maasberg2016}, an assessment approach for malware threat levels is
introduced, based on scores and weighting factors. Article \cite{Woods2015} refers to a data mining approach using similarity metrics to identify statistical
relations in shared information. In \cite{Adams2011} is described a data-driven approach to evaluate and visualize mixed content from news and social media content based on emotive facets to give value to content.

A common approach in intrusion detection for example, to refer to threshold-based methods for triggering alarms. A lot work has already been done in this
direction and a lot of surveys were published, which highlight the most
common works \cite{Mitchell}\cite{Prasad}. Threshold-based event detection
has shown to be a reliable approach in statistical, game-theoretical and data
mining evaluation techniques \cite{Barford}\cite{Nikulin}
\cite{Ghafouri} \cite{blinc}.

Beside the information that can already be analyzed for further purpose,
is the data itself, as for example the used IP-addresses, protocols,
timestamps, etc. The analysis and interpretation of these vectors also
represents a significant part of the sightings approach. One major source of
information can be netflow \cite{sommer} information for example, which can
be combined to hash functions and bloom filters, aggregation methods and
other reduction techniques for evaluation purpose
\cite{blinc}\cite{garcia}\cite{Mun}\cite{Waldvogel}\cite{Nie} \cite{Sahni}.

\section{Background information on MISP}
\label{misp-tool}

In the following paragraphs, we will shortly introduce the MISP tool and highlight
its main functionalities in order to make the sightings approach
more understandable and to illustrate, how this approach can be used in this
context. For the detailed description of the design and implementation of
MISP, we refer the reader to \cite{misp-article}.

\subsection*{General description of MISP}
 \begin{figure}[h]
\centering
\includegraphics[scale=0.28]{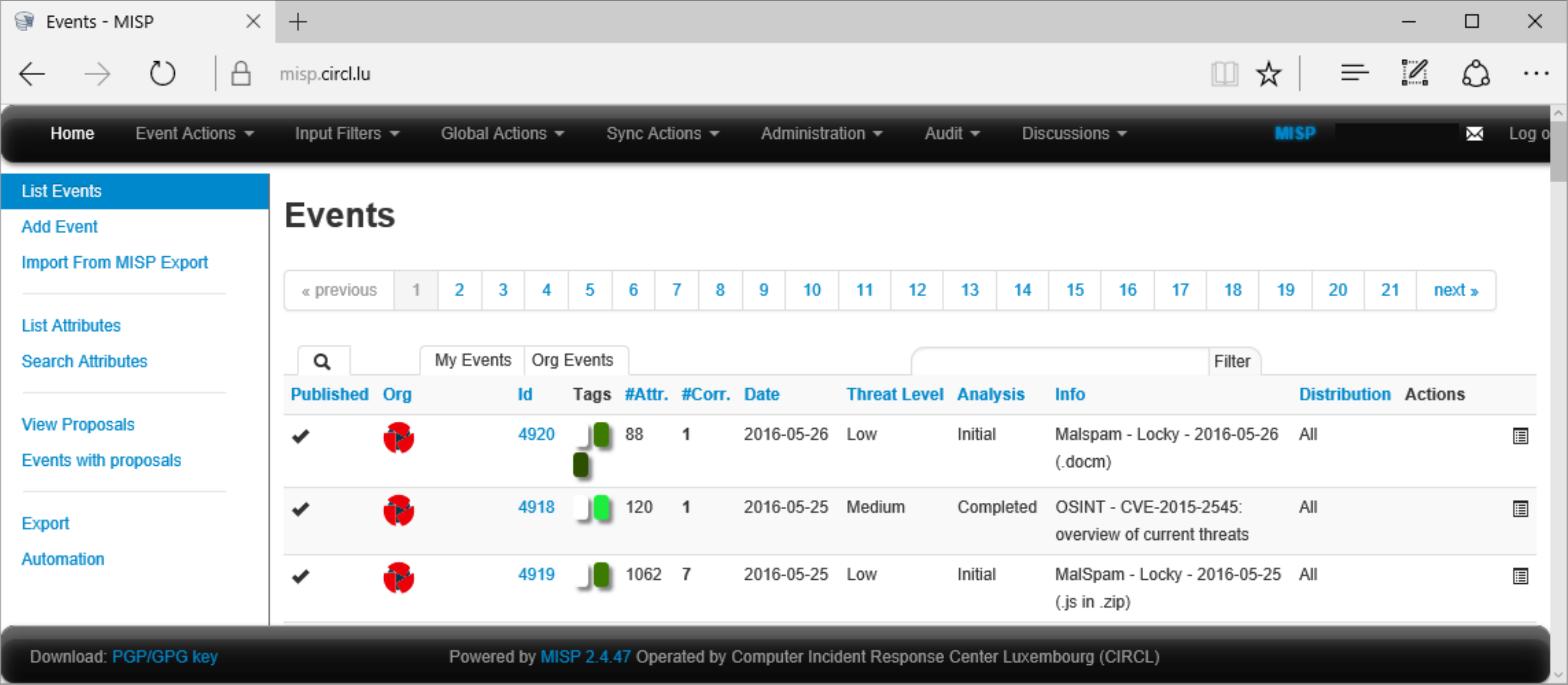}
\caption{The landing page in the MISP Interface}
\label{misp-interface}
\end{figure}

MISP is an open-source threat information sharing platform, where users from
the communities can share all kind of threats, especially all kinds of
indicators of compromise, but also others such as financial indicators as
for example  bank accounts of money mules, which were abused for money
laundering activities. The data model implemented in MISP for sharing
information is rather simple. The user can decide on the granularity of
information he wants to disclose in MISP and on the same time can also set
the sharing level (for example in his organisation only, for a specific
community only, for all sharing groups) for his information. In order to
familiarize the reader with the terminology of MISP, we present some terms
to ease the understanding of the paper.

MISP is designed to be peer to peer, where multiple instances can
exchange information with each other. The synchronization protocol in
MISP resulted from a trial-and-error approach, where the main criteria were
efficiency, accuracy and scalability. The resulting algorithm is a pull,
push and cherry-pick technique.

A shared information in MISP is called an event. An event is composed of a
list of attributes, also called attributes (destination IP addresses, file
hashes). Currently 140 types are available in MISP software. An attribute
is identified with the tuple (category, type, value). An event is also
linked with contextual information such as date, threat level, description,
organisation, galaxies about threat actors and others.

To simplify the handling of adding information to MISP and to avoid the
filling of a time-consuming and tedious form, a free text importer was
integrated, that allows users to copy and paste raw data into a single field.
This text is then analyzed by an heuristics-based algorithm to extract attributes which can be validated by the user. For the filtering of events, it is referred to
taxonomies, which will be illustrated a bit more in detail.

\subsection*{Taxonomies}

The MISP taxonomies are described in \cite{Dulaunoy}, and were
introduced for facilitating the description of IoCs and other relevant
information. It can be described as a classification scheme having its own
vocabulary.

A taxonomy in MISP is based on the machine-tag approach with
triple-tags for representing semantic information, as for example used on
Flickr for the geolocation of pictures \cite{straup2007machine}. The
triple-tag syntax is a simple expression that has a namespace, a predicate
and a value, as shown  in the following example: \{\textit{flora} :
\textit{flower} = \textit{'lily'}\}, this means that \textit{flora} is the
namespace, \textit{flower} is the predicate and 'lily' the value.

The public repository\footnote{https://github.com/MISP/misp-taxonomies} with MISP taxonomies includes 47 different taxonomies for the domains of law
enforcement, computer security incident response team (CSIRT)
classifications, intelligence and many more. In this paper, we applied
these taxonomies to the scoring model in section \ref{sighting}.

For the sake of clarity we presented MISP in this paper to introduce to
attribute scoring. For more information about MISP and its functionalities,
refer to the work of the authors of MISP in \cite{misp-article}.

\section{Data interaction methods}
\label{dint}
The typical single producer - multiple consumer paradigm frequently used in
threat intelligence platform was overthrown in MISP. Each participating
community can produce or consume information. Typical shared information
are signatures for intrusion detection systems, which can be either
exported via python API (Application Programming Interface) or with a
crafted HTTP request. An intrusion detection system can ask MISP software for the latest signatures and integrate them in its detection system. Some intrusion detection systems can be instrumented to sent a REST (REpresentational State Transfer) request towards MISP with feedback
about its detections. This feedback is called \textit{sighting}. Hence, knowledge can be gathered about the validity, freshness of an information or its impact. A typical example is an IP address of a compromised website
distributing malware or acting as command and control server. 

However, in MISP the concept of detection and sighting is not limited to network intrusion detection system, but also to host intrusion detection systems capable of detecting malicious files or more particular use cases. A common one is the detection or blocking of phone numbers in PABX systems  that are used for malicious activities such as social engineering attacks or silent or ping calls. Other examples are interaction with accounting systems that are
regularly automatically fed by MISP with recent Money mule bank accounts used in fraud cases. Hence, accountants get a warning in case they try to do a wire transfer to these numbers. A typical use case are CEO fraud attacks or intercepted and manipulated invoices. As MISP is not used
for sharing classical network intrusion signatures, the term attribute
is used in the following for designating pieces of information that are
shared or detected.

Each time a consumer receives such an attribute, the consumer can
open a discussion by entering free text with questions or comments
regarding a piece of consumed information. These comments or questions
are then propagated back trough the MISP instances.

Although, non classified threat information, for instance IP addresses
about botnets, hashes about malware, are publicly
available\footnote{\url{http://www.misp-project.org/communities/}}. For several years little research interest was there for the application of this data
as ground truth for anomaly detection or machine learning algorithms.
New attacks learned from these training sets could be suggested to the producer.


\subsection*{Sighting}

\begin{figure*}[h]
\centering
\includegraphics[scale=0.5]{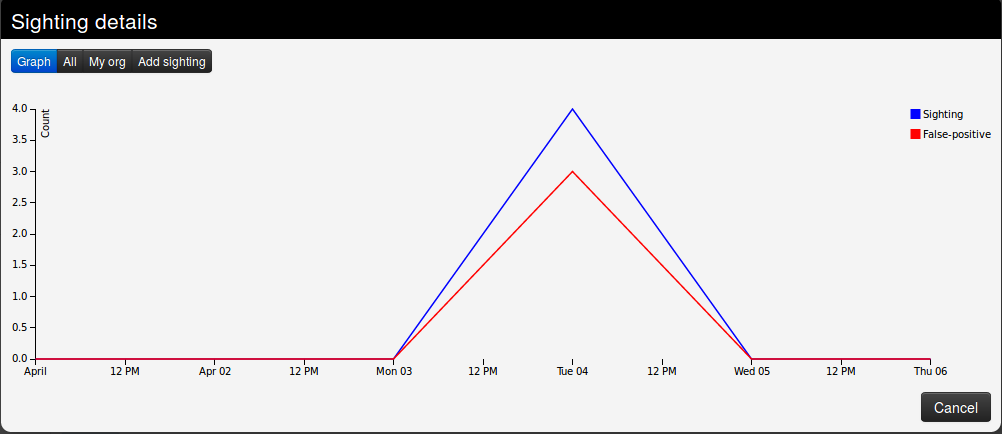}
\caption{Visual representation of sighting in MISP}
\label{sightinggraph}
\end{figure*}

MISP provides a feature called sightings where users, scripts or intrusion
detection systems can share information about a given consumed attribute.
For instance, whether they saw it or whether it is a false positive or
has expiration dates for some attributes. Sightings gives more credibility
to an attribute and can be used for prioritizing or decaying attributes.
Sightings are visually represented in MISP such as shown in figure
\ref{sightinggraph}.

\begin{figure*}[h]
\centering
\includegraphics[scale=0.5]{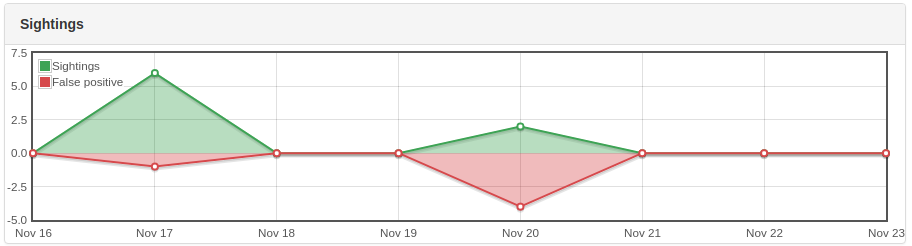}
\caption{Visual representation of the aggregation of sightings}
\label{sightingAggregraph}
\end{figure*}

Also, aggregating sightings of all attributes/objects can be useful to
detect particular security events or threats. For example, figure \ref{sightingAggregraph} is a visual representation of the occurrences of sightings and false-positive for one week. We can observe that on November 20 (2017), a lot of false positives were detected, probably indicating a spam campaign.  Also, on November 17 (2017), a larger proportion of
sightings than the normal average is represented. This can raise suspicion of
security experts or the SOC (Security Operation Center) that a more concerning threat is present. This should indicate that a deeper investigation should be performed. Sightings are valuable inputs for decaying attributes.

\section{Scoring Indicators of Compromise}
\label{sighting}
To illustrate that the handling of these attributes and the correctness of information is challenging, we will present some numbers provided from our MISP community, we operate for the private sector \cite{misppriv}. It includes
1 531 users from 761 different organisations. These users have already shared 8 101 events and 1 003 908 attributes until early December 2017.

A fact that makes the handling of these attributes even more challenging is that the objectives from each user is different and by this, having a
non-homogeneous crowd. On one hand, users want to use the data for implementing operational security, such as, performing blocking actions based on shared attributes like file hashes or IP addresses. Hence, false positives are  unwanted, since data needs to be correct and reliable. 
On the other hand, organisations want to correlate attributes and link them with additional threat actors. Hence, these organisations need a correct and reliable source of historical data.

In this section, a scoring technique for decaying attributes in MISP including
various factors such as its sightings in operational networks, its taxonomies
and the reliability of the source is presented.

The lifetime of the various available attributes are not homogeneous. For
instance, hosts of machines related to IP addresses are changed or cleaned
up, IP addresses or domain names are traded and hence get used in different
fashions over time. Hence, each attribute has its own decay function. File
hashes usually tend not to vary over time. Nevertheless, a shared file hash
can be declared as false positive over time by organizations with distinct
trust levels.

To fully evaluate the overall score of an attribute, some conditions can be
taken into consideration:
\begin{itemize}
    \item The base score of an attribute, called \texttt{base\_score}, is
           a weighting of the confidence of its source and the
          taxonomies attached to it. It is the initial value of an indicator's
          life cycle. It is also the score the indicator will be reset to
          upon a new sighting.
    \item The period of time expressed between the time an attribute
          was seen  first and the time it was seen last.
    \item The end-time of an attribute $\tau_a$ represents the time at which
          the overall score should be 0.
    \item The decay rate $\delta_a$ represents the speed at which the overall
          score is decreasing over time. It would be preferable that the decay
          speed is variable over time. To illustrate this, an example of an IP address is taken.
          The decay rate of the IP should be low for
          the first hours, but should go faster the more time passes. The first
          time activities from this IP are sighted, the better chances are
          that the threat actors are still active or are executing follow up
          operations. When this IP address is shared among a community targeted
          by the threat actor, more and more members can take measures, such as
          blocking the IP address. Hence, the attack becomes ineffective
          forcing threat actors to use other IP addresses. In case, the IP
          is given up, it could be reassigned to a legitimate customer of the
          Internet service provider leading to collateral damage due to the
          blocking actions of this IP.
\end{itemize}
MISP is a peer to peer system where people can produce and consume information
about threats in a collaborative manner. Hence, it is common that
information transits through multiple  MISP instances until it gets to
its consumer. Producers can add tags defined in taxonomies, introduced
in section \ref{misp-tool}, about their confidence or reliability of their
source about a given piece of information they are encoding. Consumers get
this information and have different levels of trust in the producers.

The $\texttt{base\_score}_a$ of an attribute is defined in equation \ref{basescore},
 $base\_score \in \left[ 0, 100 \right]$.
It represents the score of an attribute before taking
into account its decay. It is composed of its weighted applied tags and its
source confidence.

The weights of the applied taxonomies are defined at
predicate level of each taxonomy and represent its acceptance within a
community. For instance, if tags from the taxonomy with the namespace
{\tt admiralty-scale} and with the predicate {\tt source-reliability} are
hardly used, it gets a low weight. However, if within the same taxonomy tags
with the predicate {\tt information-credibility} are regularly used, it gets
a higher weight.

The \textit{source confidence}
can also be influenced by an additional parameter called $\omega_{sc}$. This
parameter takes into account more subtle trust evaluations.
For example, it could be that an organization has a good image and a
good reputation but due to some circumstances within a given time frame,
the trust in this organization is decreased. A practical example is an
organization that was compromised or taken over by the attacking party.
\begin{equation}\label{basescore}
    base\_score_{a} = weigth_{tg} \cdot tags + \omega_{sc} \cdot
    source\_confidence
\end{equation}

The $base\_score$ is defined in equation \ref{basescore} with,

\begin{itemize}
    \item $\forall weigth_{tg} \in \left[0, 100\right], \forall \omega_{sc} \in
        \left[0, 100\right], weight_{tg} + \omega_{sc} = 100$, $weight_{tg} = 100$ or
$\omega_{sc}=100$, a mean to adjust the focus either on the
$tags$ or on the $source\_confidence$. As little research on the trust
rebalancing  and trust evolution of organizations in distributed threat
        sharing is done, the  $\omega_{sc}$ parameter is set to $100-weight_{tg}$ and is considered as future work implying further research.
    \item $tags \in \left[0, 1\right]$, the score derived from the taxonomies
          is defined in equation \ref{tags}.
    \item $source\_confidence \in \left[0, 1\right]$, is the confidence
          given to the source that published the attribute.
The $source\_confidence$ parameter in equation \ref{basescore} gives a
possibility to influence the $base\_score$, which should be a number
between 0 and 100. Each source between $1$ and $N$ has its $source\_confidence$
level. In case a source is fully trusted the $source\_confidence$ is set to 1.
If there is no trust, the source level is set to 0. The user could also set
intermediate values, which could give an estimate on how reliable the
source is. The learning of the confidence of a source based on its produced
information over time is subject to future research.
\end{itemize}

The tags parameter in equation \ref{basescore} is derived from the taxonomies
a producer can attach to its piece of information.
A few used taxonomies allow to add confidence or reliability on a produced
information. The tags in the taxonomies can be attached to each individual
encoded attribute.  The taxonomies  available in MISP platforms permit
to derive confidence and reliability are defined in table \ref{confidence_table}.
\newpage
These taxonomies are: 
MISP machine tag\footnote{https://github.com/MISP/misp-taxonomies/blob/master/misp/machinetag.json},
admirality scale\footnote{https://github.com/MISP/misp-taxonomies/tree/master/admiralty-scale},
OSINT\footnote{https://github.com/MISP/misp-taxonomies/blob/master/osint/machinetag.json} and estimative language\footnote{https://github.com/MISP/misp-taxonomies/blob/master/estimative-language/machinetag.json}.
As these taxonomies are already used
by large MISP communities, a scoring model should be derived from these ones
instead of suggesting new ones, as it is unknown if the new ones will be used by the communities.

The MISP taxonomy includes a confidence level that is '\textit{Confidence cannot be evaluated}'. This special confidence level is not mapped to a numerical value. One possibility is to introduce the concept of '\textit{undefined}'.

Once a value is undefined, the $base\_score$ cannot be computed and becomes undefined. At the end, the overall score would be undefined and by this, cancel other scoring factors defined in tags. Hence, when the confidence level is '\textit{Confidence cannot be evaluated}', it will be ignored.

\begin{center}
\begin{table}
\begin{tabular}{|l|l|l|l|}
    \hline
    Description & Value(s) & Description & Value(s)\\
    \hline
    \multicolumn{2}{|c|}{Misp}&\multicolumn{2}{|c|}{OSINT}\\
    \hline
    Completely confident  & 100 &Certain & 100 \\
    Usually confident & 75& Almost certain & 93\\
    Fairly confident & 50 & Probable&75\\
    Rarely confident & 25 & Chances about even& 50\\
    Unconfident & 0& Probably not&30\\
    &&Impossibility & 0\\
    \hline
\end{tabular}
\vspace{0.3cm}
    \caption{MISP and OSINT taxonomies}
    \label{confidence_table}
\end{table}
\end{center}
\vspace{-0.9cm}
The score derived from the taxonomies is defined in equation \ref{tags}, where
$G$ is the number of defined taxonomy groups and $T$ the number of used
taxonomy per group. The weights are defined at predicate level in the taxonomies and should be integer numbers between 0 and 100.

\begin{equation}\label{tags}
    tags = \frac{\sum_{j=1}^{j=G} \sum_{i=1}^{i=T} taxonomy_{i} * weight_{i}}{{\sum_{j=1}^{j=G} \sum_{i=1}^{i=T} 100 \cdot weight_{i}}}
\end{equation}

The idea is to decrease the \texttt{base\_score} over time. When it reaches zero,
the related indicator can be discarded.
A first idea to express the overall score could be
to use  equation \ref{score1}.

\begin{equation}\label{score1}
    \texttt{score}_a = \texttt{base\_score}_a - \delta_a (T_{t} - T_{t-1})
\end{equation}
where,
\begin{itemize}
    \item \texttt{base\_score}$_a$ $\in \left [0, 100\right ]$ is described in equation \ref{basescore}.
    \item $\delta_a$ $\in \left [0, +\infty \right.$ represents the decay rate,
          or expressed as the speed at which the score of an attribute
          decreases over time.
    \item $T_{t}$ and $T_{t-1}$ are timestamps. $T_{t}$
          represents the current time and $T_{t-1}$ represents the last time
          this attribute received a sighting. Note that $T_{t} > T_{t-1}$.
\end{itemize}

\begin{figure}
\centering
    \includegraphics[scale=0.4]{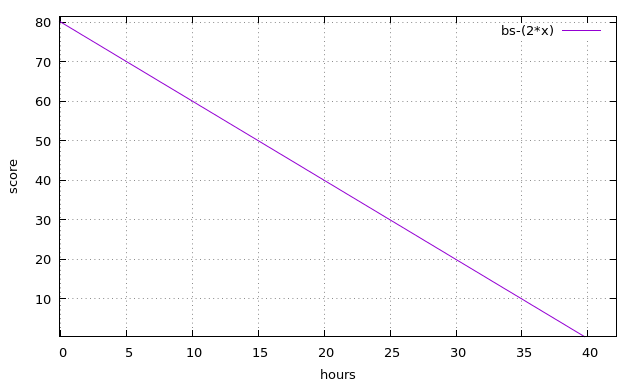}
    \caption{$\texttt{score}_a = \texttt{base\_score}_a -
              \delta_a (T_{t} - T_{t-1})$. The decay of the score is constant.}
    \label{fig:score1}
\end{figure}

Figure \ref{fig:score1} shows the decay of the score of an attribute $a$ with
a \texttt{base\_score}$_a$ of $80$ and a decay rate $\delta_a$ of $2$.

An evaliatoin of the parameters shows that neither the end-time nor the
variable decay rate can be controlled.
Indeed, by fixing the decay rate, the end-time cannot be specified for the
score of an attribute. In the same mind, even if the decay rate is controlled by the constant $\delta_a$, the decay is fixed over time.

To address the latter point, an exponential degression could be considered as
shown in equation \ref{score2}.

\begin{equation}\label{score2}
    \texttt{score}_a = \texttt{base\_score}_a \cdot e^{-\delta_a \cdot t}
\end{equation}

\begin{figure}
\centering
    \includegraphics[scale=0.4]{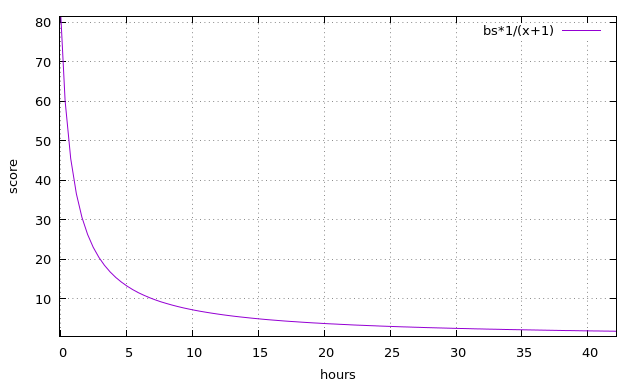}
    \caption{$\texttt{score}_a = \texttt{base\_score}_a
\cdot e^{-\delta_a \cdot t}$. The decay of the score follows an
exponential degression.}
    \label{fig:score2}
\end{figure}

In this case a variable decay rate can be used. The slope in figure \ref{fig:score2} is high at the beginning and lower as time passes. However, the decay rate cannot be significantly influenced. This expression cannot be used to have a slow decay at the beginning followed by a rapid degression. A behaviour that can, for example, be found in dynamic IP address allocation by threat actors as previously described in this section.
Moreover, the time at which the overall score of the attribute should be 0 is entirely defined by the decay rate. So, manipulating the slope as well as the end-time at the same time is still not possible.
Furthermore, it can be  observed that the choice of the parameter $\delta_a$ will essentially range between 0 and 1 due to the tendency of the exponential
degression to rapidly becomes assymptotic to 0.

The final score is defined in equation \ref{score3}, capturing the conditions
stated previously.

\begin{equation}\label{score3}
    \texttt{score}_a = \texttt{base\_score}_a \cdot \left(1 - \left(\frac{t}{\tau_a}\right )^{\frac{1}{\delta_a}}\right )
\end{equation}
with
\begin{itemize}
    \item $\delta_a \in \left ] 0, +\infty \right.$, the decay speed.
    \item $\tau_a \in \left ] 0, +\infty \right.$, the end-time or time needed such that \texttt{score}$_a = 0$. The end-time can be told by an expiration sighting, where an organization knows when an indicator will be expired. An example is the grace time: an Internet service provider gives a grace time to customers to fix their machine until disconnecting them or law enforcement agencies seizing the equipments. It can also be derived from existing regular sightings, where organizations provided data about sightings from the past.
    \item $t = T_t - T_{t-1}$, is an integer $>0$
\end{itemize}

\begin{figure}
\centering
    \includegraphics[scale=0.4]{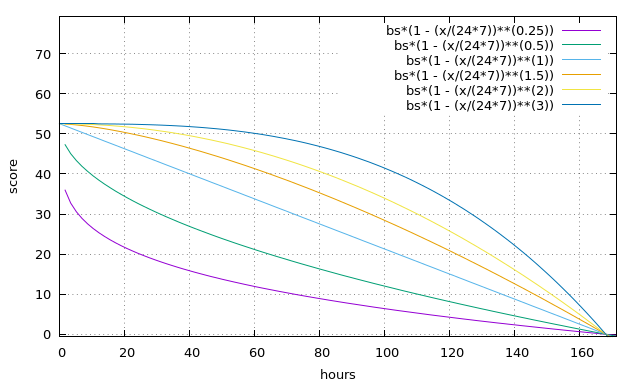}
    \caption{$\texttt{score}_a = \texttt{base\_score} \cdot \left(1 - \left(\frac{t}{\tau_a}\right )^{\frac{1}{\delta_a}}\right )$ for a fixed $\tau_a$ of 7 days}
    \label{fig:score3}
\end{figure}

This polynomial function has two advantages over the exponential one.
First, the end-time $\tau_a$ can be easily controlled.
direction of the slope's cavity. We can have fast degression at the beginning
can be obtained followed by a slow degression along with the complete opposite. An example for a different decay rate $\delta_a$ can be seen in figure \ref{fig:score3}.
It can be seen that the greater $\delta_a$ is, the faster the overall score
decreases at the beginning. The closer $\delta_a$ is to zero, the slower the overall score will decrease at the start.
The score is 0 for all decay rate for the specified $\tau_a$.

\begin{figure}
\centering
    \includegraphics[scale=0.15]{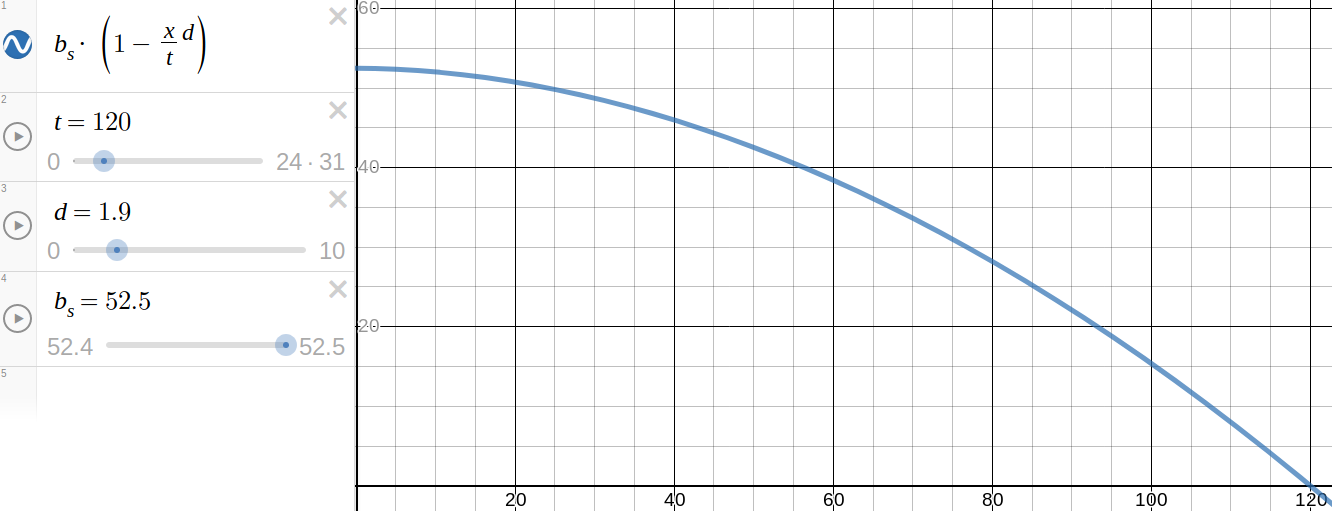}
    \caption{Web interface letting the user select the parameters. {\tiny source: www.desmos.com}}
    \label{fig:paramTuning}
\end{figure}

The parameters can easily be fine tuned, which is an additional advantage.
A value can be set for each type of attribute by performing a
statistical analysis on an existing dataset; an example of such derivation is presented in section \ref{sec:evaluation}; or users could set their
own values via a dedicated interface. An idea on how this could be done is given in figure \ref{fig:paramTuning}.

\begin{figure}
\centering
    \includegraphics[scale=0.4]{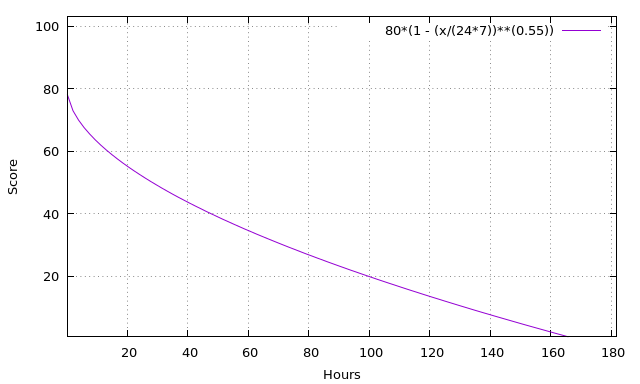}
    \caption{$\texttt{score}_a = 80 \cdot \left(1 - \left(\frac{t}{7\cdot 24}\right )^{\frac{1}{0.55}}\right )$ - It can be seen a rapid decrease of the score at the beginning. The score is halved after 48 hours.}
    \label{fig:ex1}
\end{figure}

Two examples are shown on how the score in equation \ref{score3} can be used.
The first example is an attribute for
a compromised IP address being part of a botnet. The attribute of a shared
event in MISP belongs to the category {\tt Network activity} with its type
{\tt ip-dest}, meaning the destination IP address of a compromised webserver
hosting an exploitkit distributing malware. Some organizations spotted it
and started to share information about it. Abuse teams  are informed to cleanup the compromised systems. The IP address is encoded in publicly available blacklists. The threat actors might notice the detection too and start to move their exploitkit to another webserver. If we assume that the Internet service provider gives a customer 1 week time to fix the webserver. If it is not fixed within this time frame, the IP of the webserver will be null-routed,
meaning that it will not be accessible any more. Hence, $\tau_a = 7\cdot 24$ hours. Under the hypothesis that the typical blacklists take 48 hours to be applied in proxy servers or browsers, the overall score should be halved after 2 days. Hence, $\delta_a = 0.55$.
Finally, if the base score of the attribute is calculated to be \texttt{base\_score}$ = 80$ (based on the taxonomies and source confidence),
equation \ref{score3} becomes:

$$\texttt{score}_a = 80 \cdot \left(1 - \left(\frac{t}{7\cdot 24}\right )^{\frac{1}{0.55}}\right )$$
where $t$ is the time between now and the last \textit{sighting}, expressed
in hours. A plot of the decay is represented in figure \ref{fig:ex1}.

\begin{figure}
\centering
    \includegraphics[scale=0.4]{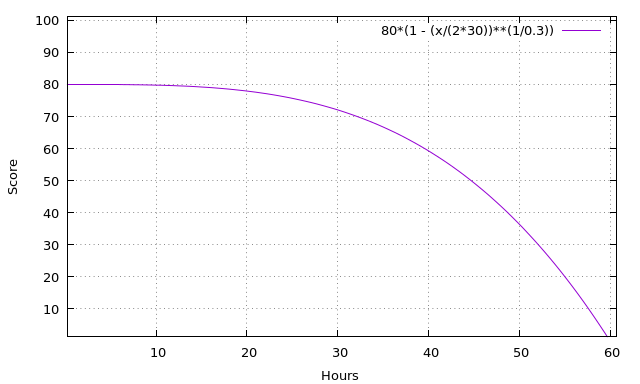}
    \caption{$\texttt{score}_a = 80 \cdot \left(1 - \left(\frac{t}{2\cdot 30}\right )^{\frac{1}{0.3}}\right )$ -  A really slow decrease at the beginning and a rush towards zero at the end can be observed. The overall score is halved only after 48 hours.}
    \label{fig:ex2}
\end{figure}

The second example is the hash of a malware. In this scenario, a file-hash is not as volatile as an IP. It could be considered
that the attribute will not have any value after 2 months, with a rather slow
decay. We can now set $\tau_a = 2\cdot 30$ days and $\delta_a = 0.3$.
It is also supposed that the base score is the same as the previous example:
\texttt{base\_score}$= 80$. We have:
$$\texttt{score}_a = 80 \cdot \left(1 - \left(\frac{t}{2\cdot 30}\right )^{\frac{1}{0.3}}\right )$$
and the resulting plot can be seen in figure \ref{fig:ex2}.

\section{Experimental evaluation: tuning $\tau$ and $\delta$ using a phishing dataset \label{sec:evaluation}}
In this section, we show how to fine-tune the two parameters of the model using basic statistical analysis
on a real dataset.

The original dataset consists in attributes of type URL related to phishing campaigns,
available at SOME URL\footnote{\url{SOME URL}{SOME URL}}. Statistics about the dataset are shown in table \ref{table:dataset}.

\begin{figure}[h!]
    \begin{tabular}{|ll|}
        \hline
        Time spanned & May 24, 2017 $\rightarrow$ May 3, 2018\\
        Number of attribute & 583783\\
        Number of sighting & 92713361\\
        Mean ($\mu$) of sighting / attribute & 333\\
        Stdev ($\sigma$) of sighting / attribute & 1071\\
        \hline
    \end{tabular}
    \label{table:dataset}
    \caption{Statistics on the dataset used to evaluate the parameters}
\end{figure}

This dataset has been processed in order to deduce the end-time of each attributes using sightings.
For each of them, the end-time was calculated as follow:

$$ \texttt{end-time} = (t_n - t_0) + \Delta_{max} $$

Where,
\begin{itemize}
    \item $t_0$ is the time at which the attribute was seen for the first time
    \item $t_n$ is the time at which the attribute was seen for the last time
    \item $\Delta_{max}$ is the longer time elapsed between 2 sightings
\end{itemize}

\begin{figure}[!h]
    \centering
    \includegraphics[width=0.46\textwidth]{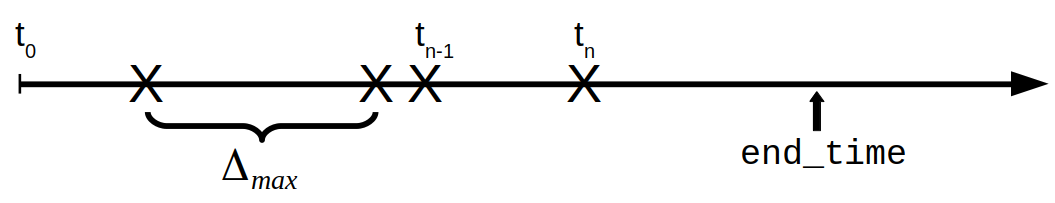}
    \caption{Timeline showing the method to compute the \texttt{end\_time}}
\end{figure}

From a CSIRT perspective, it is critical to take down (or request it) compromised servers as fast as possible.
Even if long-living URLs are important, we only consider short-lived URLs in the evaluation. Longer URLs are a special case that is out of scope for this section and so, are treated as outliers and ignored.

\begin{figure}[h]
    \centering
    \hspace{-0.05\textwidth}
    \includegraphics[width=0.52\textwidth]{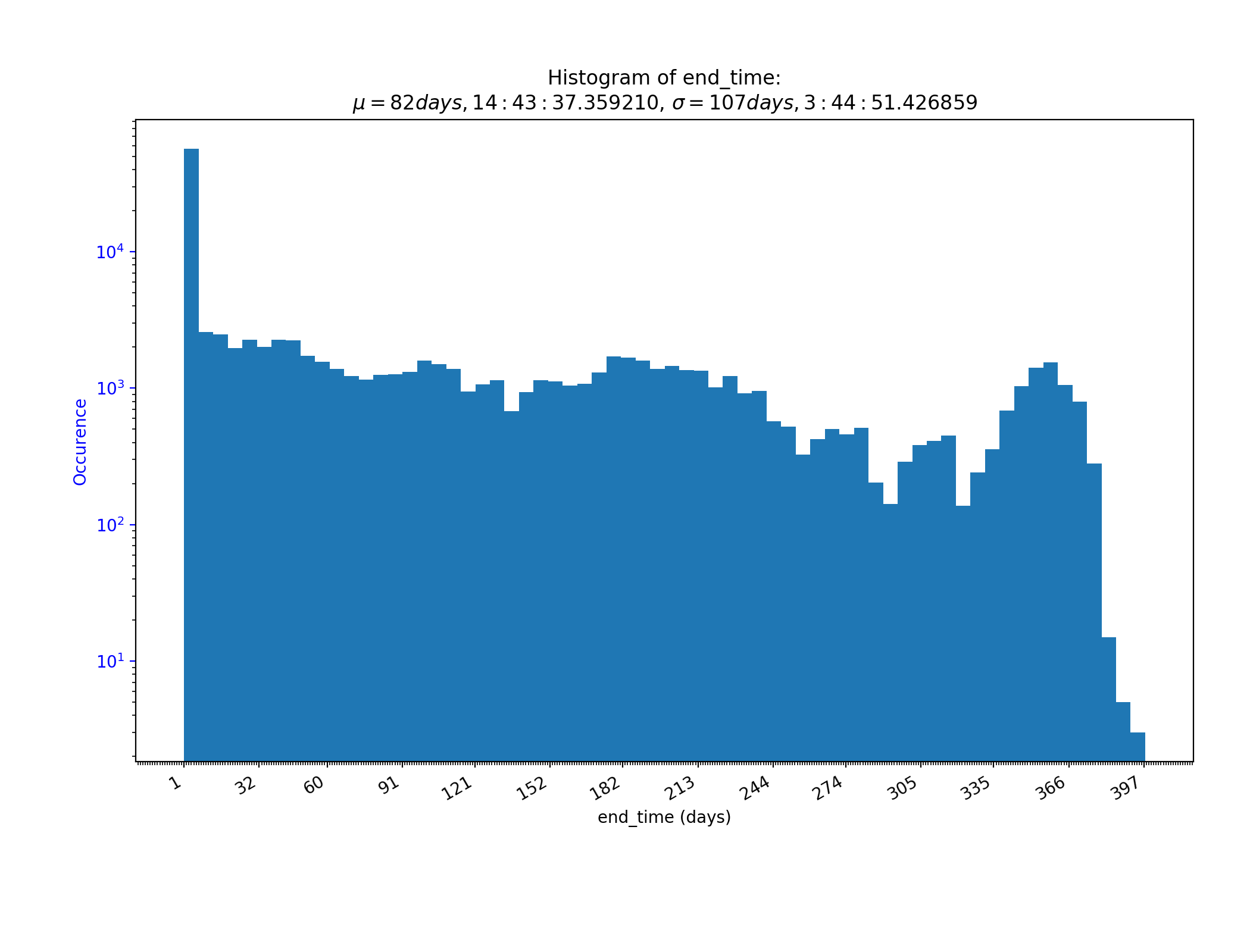}
    \label{histo1}
    \caption{Histogram of the \texttt{end\_time} for the complete dataset (log scale)}
\end{figure}

This processed dataset can be used to deduce the two parameters\footnote{In this section, the \texttt{base\_score} is assumed to be 100}. The first one we are interested in is $\tau$ (the end-time).
By looking at the histogram \ref{histo1} showing the repartition of the calculated end-time over time\footnote{Note the log scale}, we can see that the majority
of attributes are located within the first week.

\begin{figure}[h]
    \centering
    \hspace{-0.05\textwidth}
    \includegraphics[width=0.52\textwidth]{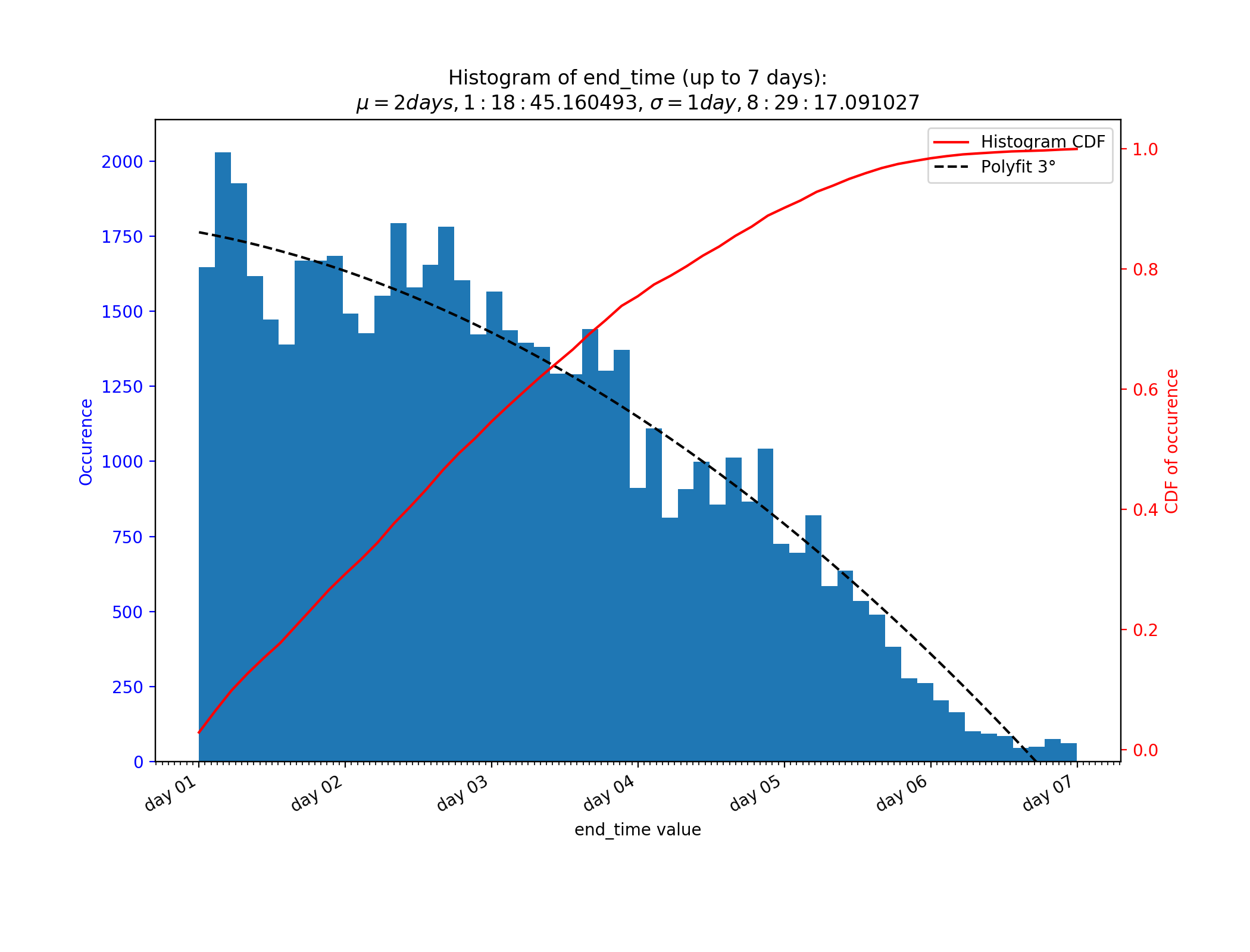}
    \label{histo2}
    \caption{Histogram of the calculated \texttt{end\_time} for the first week}
\end{figure}

By zooming in the first week we obtain the histogram \ref{histo2} where more information can be obtained.

As stated previously, the majority of the end-time is sitting in the first week.
Moreover, the CDF indicate that $\sim 90\%$ of the attributes shown on the graph falls within 5 days.
Then, it is reasonable to consider that the end-time is:
$$ \tau = 5\;days = 120 \;hours $$




The value of  the second parameter $\delta$ can be estimated by looking at the general tendency of the histogram.
It is clear that the slope cavity is directed towards the bottom, indicating an increasing speed of decay over the time.
Even before computing $\delta$, we can see that this parameter will be $>0$.

Again, by looking at the CDF, $50\%$ of the attributes for the first week are located around 3 days ($72\; hours$).
Then, it is also reasonable to say that after 3 days, half of the attribute have expired. Hence, the score at this point is 50.
We can now compute $\delta$:
\begin{align*}
    \texttt{score} &= \texttt{base\_score} \cdot \left(1 - \left(\frac{t}{\tau}\right )^{\frac{1}{\delta}}\right ) \\
    \llap{$\Leftrightarrow$\hspace{20pt}} 50 &= 100 \cdot \left(1 - \left(\frac{72}{120}\right )^{\frac{1}{\delta}}\right ) \\
    \llap{$\Leftrightarrow$\hspace{25pt}} \delta &\simeq 1.3
\end{align*}

Finally, attributes of type URL concerning phishing campaing can use these estimated parameters to be scored.

This evaluation can be used by IDS to select which rules to play on. Practically, IDS are limited on the number of entry they can charge at the same time,
and so, only a portion of attributes can be loaded as a rule. By using such instantiation of the presented model and an appropriate threshold on the score,
an IDS could support volatile URL partaining in phishing camapaign.

\begin{figure}[h]
    \centering
    \hspace{-0.05\textwidth}
    \includegraphics[width=0.52\textwidth]{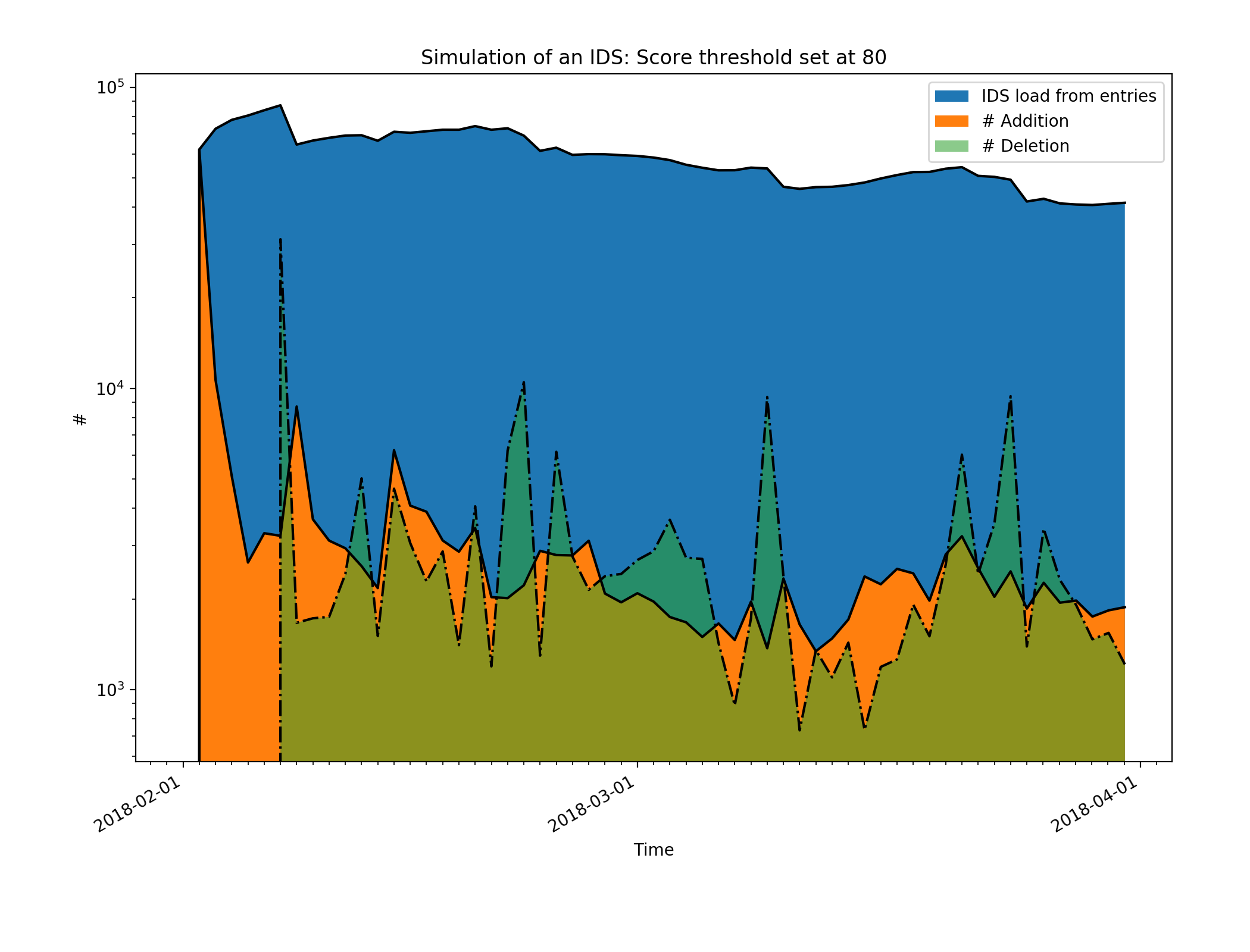}
    \label{graph:simulation}
    \caption{Simulation of an IDS supporting the model}
\end{figure}

\begin{figure}[h]
    \centering
    \includegraphics[width=0.45\textwidth]{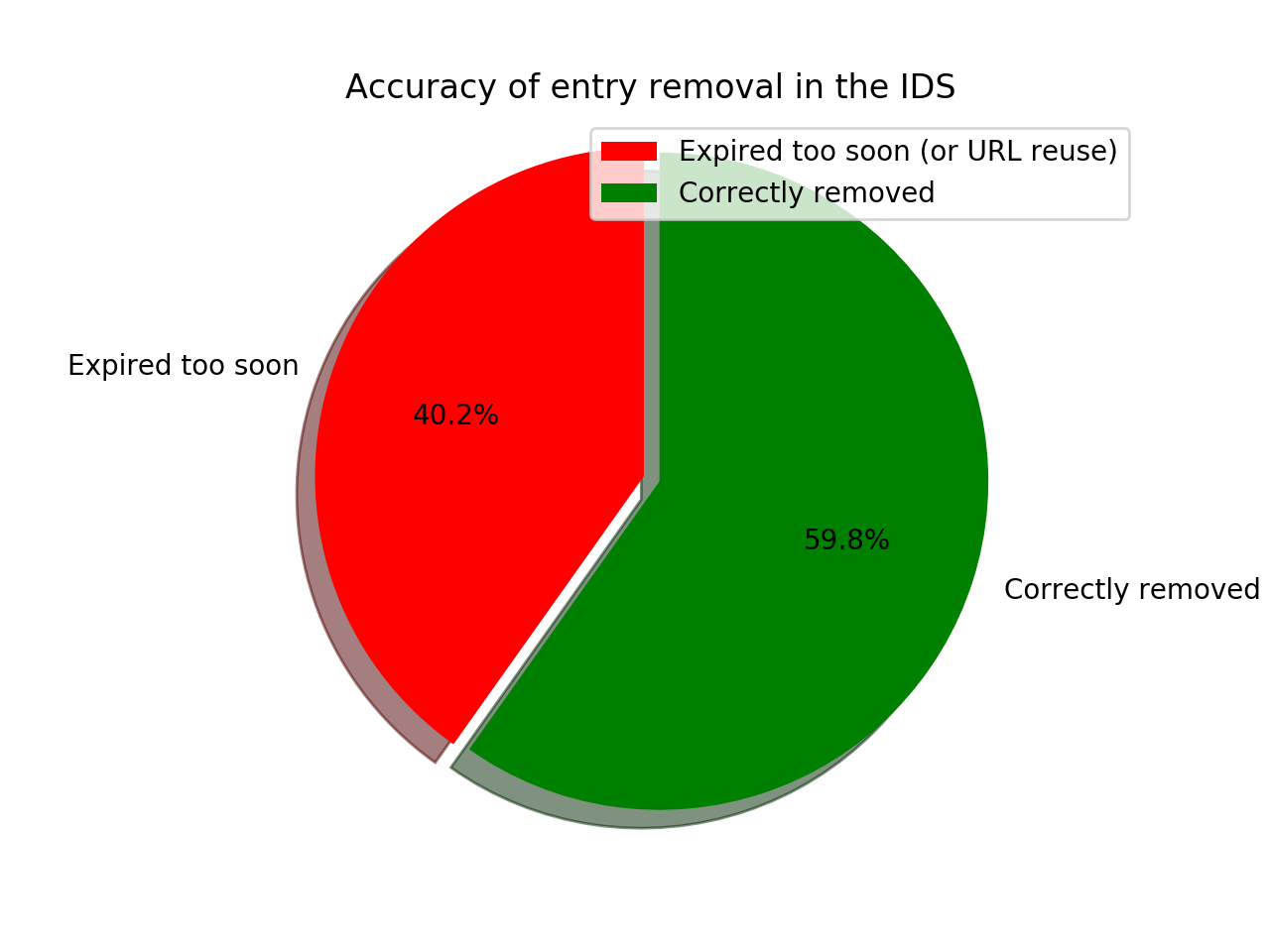}
    \label{graph:simulation2}
    \caption{Percentage of correctly vs prematurely removed entries}
\end{figure}

A simulation of this is shown in the chart \ref{graph:simulation} and \ref{graph:simulation2}.
The simulation has been done for a period of 2 months using the data from the initial dataset.
It can be seen that nearly 60\% of the entries were correctly removed from the IDS table, while 40\% of them were removed prematurily.

\section{Future Work and Conclusions}
\label{conclusions}

Information sharing has become an integrated part in the resolution of
incidents nowadays. The MISP platform is a unique tool that not only allows
the sharing of information, but also allows to contribute useful addons by the
community in a trusted environment.

In this paper, we presented early works on scoring mechanisms for attributes
that are shared within MISP. As MISP is a distributed peer to peer threat sharing
system, where each participating organization can consume or produce information, a consumer can receive information multiple hops away. Hence, a consumer has to somehow trust the producing organisation. The producing organization has some taxonomies in MISP for attaching reliability or credibility
to the attribute it is sharing. In this paper a base score is defined to combine these different trust aspects. Due to the different life times of attributes, a scoring method is presented, taking into account standard data interaction methods in MISP.

In distributed threat sharing models such as MISP, trust is an essential factor.
In future research activities various models for the $source\_confidence$
will be evaluated. A potential research track is the evaluation and application of
machine learning techniques. However, for evaluating these, the simple model
described in \ref{sighting} must be operational such that a significant dataset about sharing behaviours can be collected. Another research track, which might need less operational data is the exploration of game theoretical models in
context of distributed information sharing. MISP was growing organically,
where only a few MISP instances were interconnected and where the members of the various communities knew and trusted each other. However, today, it is
not uncommon that a piece of information transits through a couple of
MISP instances and only little is known about the producer or even the source of information.
Hence, it could be that some adversaries share fake information for harming
organizations or disrupting sharing communities. Detailed adversary models in
the context of peer to peer threat sharing might be useful to be studied.

\section{Acknowledgment}

This work is co-financed by the European Union under the CEF grant 2016-LU-IA-0098.

%




\begin{thebibliography}{1}


\bibitem{Adams2011}
B.~Adams, D.Q.~Phung, and S.~Venkatesh.
\newblock Eventscapes: visualizing events over time with emotive facets.
\newblock In {\em Proceedings of the 19th International Conference on Multimedia 2011}, pages 1477--1480,Scottsdale, AZ, USA, 2011.ACM.

\bibitem{Barford}
P.~Barford, J.~Kline, D.~Plonka and A.~Ron.
\newblock A signal analysis of network traffic anomalies. ACM Sigcomm IMW, 2002.


\bibitem{Brown2015}
S.~Brown, J.~Gommers, and O.~Serrano.
\newblock From cyber security information sharing to threat management.
\newblock In {\em Proceedings of the 2Nd ACM Workshop on Information Sharing and Collaborative Security}, WISCS '15, pages 43--49, New York, NY, USA, 2015. ACM.

\bibitem{misppriv}
CIRCL.
\newblock Misppriv.
\newblock https://misppriv.circl.lu, 2016.

\bibitem{cybox}
CybOX.
\newblock https://cyboxproject.github.io , 2017.

\bibitem{Dandurand13}
L.~Dandurand and O.~Serrano.
\newblock Towards improved cyber security information sharing.
\newblock In {\em Cyber Conflict (CyCon), 2013 5th International Conference on}, pages 1--16, 2013.

\bibitem{Dulaunoy}
A.~Dulaunoy and A.~Iklody.
\newblock MISP taxonomy format, https://tools.ietf.org/html/draft-dulaunoy-misp-taxonomy-format-00, 2016.

\bibitem{garcia}
S.~Garcia-Jimenez, E.~Magana, M.~Izal and D.~Moratoand.
\newblock IP addresses distribution in Internet and its application on reduction methods for IP alias resolution.
\newblock 34th Conference on Local Computer Networks. LCN 2009. IEEE.

\bibitem{Ghafouri}
A.~Ghafouri, W.~Abbas, A.~Lazka, Y.~Vorobeychik and X.~Koutsoukos.
\newblock Optimal Thresholds for Anomaly-Based Intrusion Detection in Dynamical Environments.
\newblock ArXiv e-prints 1606.06707, 2016.

\bibitem{Enisa}
U.~Helmbrecht, S.~Purser, G.~Cooper, D.~Ikonomou, L.~Marinos, E.~Ouzounis, M.~Thorbrugge, A.~Mitrakas, and S.~Capogrossi.
\newblock Cybersecurity cooperation: Defending the digital frontline.
\newblock Technical report, ENISA, October 2013.


\bibitem{Haass2015}
J.~C. Haass, G.-J. Ahn, and F.~Grimmelmann.
\newblock Actra: A case study for threat information sharing.
\newblock In {\em Proceedings of the 2Nd ACM Workshop on Information Sharing and Collaborative Security}, WISCS '15, pages 23--26, New York, NY, USA, 2015. ACM.

\bibitem{Prasad}
V.~Jyothsna and R.~Prasad.
\newblock A Review of Anomaly based Intrusion Detection Systems.
\newblock International Journal of Computer Applications, 2011.



\bibitem{blinc}
T..~Karagiannis, K.~Papagiannaki and M.~ Faloutsos
\newblock BLINC: Multilevel Traffic Classification in the Dark.
\newblock ACM SIGCOMM05, Philadelphia, Pennsylvania, USA, 2005.


\bibitem{Maasberg2016}
M.~Maasberg, M.~Ko, and N.~L. Beebe.
\newblock Exploring a systematic approach to malware threat assessment.
\newblock In {\em 49th Hawaii International Conference on System Sciences (HICSS)}, pages 5517--5526, 2016.

\bibitem{misp-book}
MISP~Contributors.
\newblock User guide of misp malware information sharing platform, a threat
  sharing platform.
\newblock https://www.circl.lu/doc/misp/book.pdf, 2017.

\bibitem{Mitchell}
R.~Mitchell and I.R.~Chen.
\newblock A Survey of Intrusion Detection Techniques for Cyber-physical Systems.
\newblock ACM Comput. Surv., doi 10.1145/2542049, 2014.

\bibitem{Mun}
J.H.~Mun and H.~Lim.
\newblock New Approach for Efficient IP Address Lookup Using a Bloom Filter in Trie-Based Algorithms.
\newblock in IEEE Transactions on Computers, vol. 65, no. 5, pp. 1558-1565, May 1 2016.

\bibitem{Nikulin}
V.~Nikulin.
\newblock Threshold-based clustering for intrusion detection systems.
\newblock Society of Photo-Optical Instrumentation Engineers (SPIE) Conference Series,  doi 10.1117/12.665326, 2006.


\bibitem{sommer}
R.~Sommer.
\newblock NetFlow: Information loss or win?
\newblock In {\em Proceedings of the 2nd ACM SIGCOMM Workshop on Internet measurement}, pp. 173-174, 2002.

\bibitem{straup2007machine}
A.~Straup~Cope.
\newblock Machine tags. flickr.
\newblock https://www.flickr.com/groups/api/discuss/72157594497877875/, 2007.

\bibitem{Murdoch2015}
S.~Murdoch and N.~Leaver.
\newblock Anonymity vs. trust in cyber-security collaboration.
\newblock In {\em Proceedings of the 2Nd ACM Workshop on Information Sharing and Collaborative Security}, WISCS '15, pages 27--29, New York, NY, USA, 2015. ACM.

\bibitem{Nie}
X.~Nie, D.J.~Wilson, J.~Cornet, D.~Damm and Y.~Zhao.
\newblock IP address lookup using a dynamic hash function.
\newblock Canadian Conference on Electrical and Computer Engineering, 2005.

\bibitem{Stix}
S.~Barnum.
\newblock Standardizing cyber threat intelligence information with the structured threat information expression (stix).
\newblock Technical report, MITRE Corporation, 2012.

\bibitem{Sahni}
S.~Sahni, and K.~Kim.
\newblock  Efficient Construction of Multibit Tries for IP Lookup.
\newblock IEEE/ACM Transactions on Networking. Vol. 11, Aug 2003. 


\bibitem{taxii}
TAXII~project.
\newblock https://oasis-open.github.io/cti-documentation/ , 2017.

\bibitem{misp-article}
C.~Wagner, A.~Dulaunoy, G.~Wagener and A.~Iklody.
\newblock MISP: The Design and Implementation of a Collaborative Threat Intelligence Sharing Platform.
\newblock Proceedings of the 2016 ACM on Workshop on Information Sharing and Collaborative Security (WISCS'16), pages 49--56, 2016.
\newblock doi: 10.1145/2994539.2994542.

\bibitem{Waldvogel}
M.~Waldvogel, G.~Varghese J.~Turner and B.~Plattnery.
\newblock Scalable High Speed IP Routing Lookups.
\newblock In Proceedings of IEEE ACM SIGCOMM 97 Cannes, France, pp.25-36,
1997. 

\bibitem{Woods2015}
B.~Woods, S.~Perl, and B.~Lindauer.
\newblock Data mining for efficient collaborative information discovery.
\newblock In {\em Proceedings of the 2Nd ACM Workshop on Information Sharing and Collaborative Security}, WISCS '15, pages 3--12, New York, NY, USA,2015. ACM.


















\end{thebibliography}
%

\end{document}